\documentclass[]{article}
\usepackage{latexsym}
\usepackage{epsfig}

\begin{document}
\baselineskip=5mm
\quad

\vspace{28pt} \centerline{\Large \bf Hydrodynamic Evolution of
Spherical Fireball}

\vspace{4pt} \centerline{\Large \bf In Relativistic Heavy Ion
Collisions}

\vspace{32pt} \centerline{\large Hong Miao$^1$, Zhongbiao Ma$^1$,
Chongshou Gao$^{1,2}$}

\vspace{16pt} \centerline{\large $^1$Department of Physics,
Peking University, Beijing 100871, China}

\vspace{4pt} \centerline{\large $^2$Institute of Theoretical
Physics, Academia Sinica, Beijing 100080, China}

\vspace{16pt}
\begin{abstract}

Evolution process could be calculated from the relativistic
hydrodynamic equation with certain estimated initial conditions
about a single spherical fireball here. So one could estimate a
kind of initial condition qualitatively with a possible energy
density about $\epsilon_0\approx 1.9$ $GeV/fm^3$, based on this
process to fit the experimental data at thermal freeze-out. The
evolution from a cylindrical fireball will be discussed simply in
a later chapter.
\vspace*{8pt} \\
{\bf PACS number(s)\/}: 12.38.Mh 25.75.-q
\\{\bf Key words\/}: QGP, plasma, hydrodynamic, evolution, fireball
\end{abstract}
\vspace{12pt}

\leftline{ \bf 1. Introduction }

\vspace{8pt}

It has been suggested that in relativistic heavy ion collisions
the QGP state would be formed, in which quarks and gluons are
free to roam within the volume of the fireball created by the
collision\cite {sign:sign1}\cite {sign:sign2}\cite
{sign:sign3}\cite {sign:sign4}. Studying the hydrodynamic
evolution of the fireball would be helpful to learn the QGP and
the phase transition from QGP to hadron gas or whether it has
happened\cite {sign:sign5}. The evolution of a fireball with
certain simple (energy density) distributions in central incident
is computed to give some coarse estimations of the initial energy
density and other information that may be the signatures of the
QGP\cite {sign:sign6}.

In our calculations, some conditions are presumed. (1) The
evolution could be described as a quasi-static process, as well
as the local thermal equilibrium is formed. Every small mass
region could be described with the mean energy density, pressure
and c.m. velocity. (2) Ideal gas is provided, therefore
$p=\frac{1}{3}\epsilon$. (3) The fireball given has a spherical
shape and there are no revolving fluid movements in central
incident, all movements of the relativistic fluid are radial. $
p,$ $\epsilon $ are also radial fields.

In fact, the original shape is not known exactly, especially there
is a finite time during the collision proceed. The fireball under
building also evolves at the same time. For Na49, it is about 1.5
$fm/c$. Therefore, spherical model is used here, for the
facilities of computing and also due to our confidence that the
difference could be accepted comparing to uncertainties by other
causes in the calculations. Cylindrical expansion from a flat
fireball will be discussed in Chapter 5.

With these conditions provided, one can use the relativistic
hydrodynamic equation 

\begin{equation}
 \frac{\partial T_{\mu\nu}}{\partial x_\mu}=0,
\end{equation}
where

\begin{equation}
T_{\mu\nu}=(\epsilon+p)u_{\mu}u_\nu-g_{\mu\nu}p,
\end{equation}
to calculate it.

\vspace{12pt}

\leftline{ \bf 2. Hydrodynamic Evolution }

\vspace{8pt}

In order to describe the global fireball properly and easily,
variables in the form of spherical coordinate
($r,\theta,\varphi$) are used while still keeping the hydrodynamic
equation in Minkowski from. Because the fluid velocity is always
radical as been mentioned, one has (4-velocity)
$u_\theta=u_\varphi=0$, and
\begin{eqnarray*}
u_x&=&u_r\sin\theta\cos\varphi  ,\\
u_y&=&u_r\sin\theta\sin\varphi  ,\\
u_z&=&u_r\cos\theta.
\end{eqnarray*}

Now Eq (1) can be transformed by this way
\begin{eqnarray}
\nonumber  \frac{\partial T_{\mu\nu}}{\partial
x_\mu}&=&\frac{\partial T_{0\nu}}{\partial t}+\matrix{{ _3}\cr
\sum
\cr{^{i=1}}}\frac{\partial T_{i\nu}}{\partial x_i} \\
&=&\frac{\partial T_{0\nu}}{\partial t}+\matrix{{_3}\cr \sum \cr
^{i=1}}(\frac{\partial T_{i\nu}}{\partial r}\frac{\partial
r}{\partial x_i}+\frac{\partial T_{i\nu}}{\partial
\theta}\frac{\partial \theta}{\partial x_i}+\frac{\partial
T_{i\nu}}{\partial \varphi}\frac{\partial \varphi}{\partial x_i}).
\end{eqnarray}

Based on the relations above one gets that, when $\nu=1,2,3$,
equations are same
\begin{equation}
(\epsilon+p)\left[\frac{\partial u^2 _r}{\partial
r}+\frac{\partial (u_ru_0)}{\partial t}+2\frac{u^2
_r}{r}\right]+u^2 _r\frac{\partial\ (\epsilon+p)}{\partial r}
+u_ru_0\frac{\partial(\epsilon+p)}{\partial t}+\frac{\partial
p}{\partial r}=0,
\end{equation}
when $\nu=0$,
\begin{equation}
(\epsilon+p)\left[\frac{\partial (u_ru_0)}{\partial
r}+\frac{\partial u^2 _0}{\partial
t}+2\frac{u_ru_0}{r}\right]+u_ru_0\frac{\partial\
(\epsilon+p)}{\partial
r}+u_0^2\frac{\partial(\epsilon+p)}{\partial t}-\frac{\partial
p}{\partial t}=0.
\end{equation}

For $p=\frac{1}{3} \epsilon$, and
\begin{equation}
u^2 _0-u^2 _r=1,
\end{equation}
(4)(5) can be simplified as
\begin{eqnarray}
\nonumber \frac{\partial\ln \epsilon}{\partial r}&=&\frac{8}{3}
\left[-u_r\frac{\partial u_r}{\partial r}-\frac{2u^2 _r+3}{2u_0}\frac{\partial u_r}{\partial t}+\frac{u^2 _r}{r}\right] ,\\
\frac{\partial\ln \epsilon}{\partial t}&=&\frac{8}{3}
\left[\frac{2u^2 _r-1}{2u_0}\frac{\partial u_r}{\partial
r}+u_r\frac{\partial u_r}{\partial t}-\frac{u_ru_0}{r}\right].
\end{eqnarray}

For $v=v_r=u_r/u_0$, one has
\begin{eqnarray}
\nonumber \frac{\partial u_r}{\partial x_\mu}&=&u^3 _0\frac{\partial v}{\partial x_\mu} ,\\
u^2 _r&=&\frac{v^2}{1-v^2},\\
\nonumber u^2 _0&=&\frac{1}{1-v^2}=\gamma^2,
\end{eqnarray}
and Eqs (7) turn to
\begin{eqnarray} \nonumber \frac{\partial
v}{\partial t}&=&-\frac{2}{3-v^2}
\left[v\frac{\partial v}{\partial r}+(1-v^2)^2(\frac{3}{8}\frac{\partial\ln \epsilon}{\partial r})-\frac{v^2(1-v^2)}{r}\right]  ,\\
\frac{\partial\ln \epsilon}{\partial t}&=&
-\frac{1}{3-v^2}\cdot\frac{8}{3}\left[\frac{3}{2}\frac{\partial
v}{\partial r}+2v(\frac{3}{8}\frac{\partial\ln \epsilon}{\partial
r})+3\frac{v}{r}\right].
\end{eqnarray}

They are non-linear partial differential equations which could
describe the expansion process of a spherical fireball. From (9)
one can find that the geometrical shape and velocity distribution
are invariant to the scale of energy density, from the form
($\partial \ln \epsilon$). This is because the equation of states
of ideal gas ($p=\frac{1}{3}\epsilon$) is used. This makes some
geometrical data independent of the initial energy density, but
only depend on its relative distribution (i.e. the value of
$\sigma$) in this model. \vspace{12pt}

\leftline{ \bf 3. Calculations }

\vspace{8pt}

Two-step Lax-Wendroff Method \cite {sign:sign7} is used here to
compute the evolution. Gaussian distribution and mean
distribution of energy density are tried as the initial
conditions to run the programmes. Gaussian condition works much
better than the other in the calculations.

In order to minimize the computational error and to raise the
stability of the programme, a small variable $\epsilon _v $ is
added as a correction to the energy density to avoid it too close
to zero
\begin{equation}
\epsilon=\epsilon_{phy}+\epsilon_v.
\end{equation}
This correction is so small that it has not a little influence on
the final results, but it helps the programme to compute the
evolution for a long time (6-10 $fm/c$) enough to freeze-out. It
could be regarded as the energy density of the base state of
physical vacuum.

The programme could output a series of data of energy density,
fluid velocity and size. The Root Mean Square (RMS) Radius
($R_{rms}$) is
\begin{equation}
R_{rms} ^2=\frac{(\Delta r)^5\pi}{E_s}[\frac{\epsilon
(0)}{40}+4\sum \epsilon(n)n^4],
\end{equation}
where $E_s$ is the total rest energy
\[E_s=(\Delta r)^3\pi[\frac{\epsilon
(0)}{6}+4\sum \epsilon(n)n^2] ,\]
and $\Delta r$ is the stepsize about location. The relation to the
effective radius (RMS on projection to one dimension) is
\begin{eqnarray}
\sigma&=&\frac{\sqrt{3}}{3}R_{rms} .
\end{eqnarray}
It is interesting that $(E_s*R_{rms} ^2)$ looks conservative
during the fireball expands, no matter which initial conditions
it has. It should be a mirror of hydrodynamic equation
(energy-momentum coservation). After setting some acceptable
conditions on the borders, the programme could evolve about $ 12
fm/c $ stably.

\vspace{12pt}

\leftline{ \bf 4. Experimental Analysis and Evolution Results }

\vspace{8pt}

The data can be observed in the experiments are the overall time
of expansion $t$, the transverse energy distribution via
pseudorapidity $dE_T/d\eta$, the emission source size $\sigma$ and
the transverse velocity $v_T$ when the the fireball freezes out.
Initial inputs are central energy density $\epsilon_0$ and the
effective size of its distribution $\sigma_0$. Trying appropriate
$\epsilon_0$, $\sigma_0$ and a fixed $t$, can fit the other
experimental data very well.

The volume of the cross section of central colliding region in
c.m. frame can be estimated as\cite {sign:sign6}
\begin{equation}
V_0 = \frac{4 \pi }{3} R_0^3 \frac{A_1^2 + {A'}_2^2}{\sqrt{A_1^2
+ {A'}_2^2 + 2 A_1 {A'}_2 \cosh Y_L}},
\end{equation}
where $ Y_L $ is the rapidity of incident nucleus in the
laboratory frame.

The radius of this region $r_c$ is about 1.5 $fm$ for RHIC and
3.26 $fm$ for SPS 158 $GeV$ $^{208} Pb$. According to a Gaussian
distribution, this radius should be smaller than the RMS radius
and larger than the effective radius, that is
\[\sigma_0\leq r_c\leq R_{rms}.\]
So from eq (13), for SPS one has
\[1.89 fm\leq\sigma_0\leq 3.26fm.\]
Initial effective radius will be tried from this domain.

The overall time of expansion (to freeze-out) $t$ is observed
about 8 $fm/c$ near mid rapidity, decreasing slightly to 6 $fm/c$
at high rapidity, with a Gaussian radius (mean square error)
$\sigma=R_0=\Delta \tau=$3.5 $fm/c$ \cite {sign:sign8}. We set it
at $t=7.5$ $fm/c$ as the time when freeze out.

Transverse energy distribution calculated between 2$\leq\eta\leq$4
shows that there is a peak near $\eta\geq y_{c.m.}$=2.9 and
depended on the initial effective size $\sigma_0$, because
pseudorapidity is used instead of rapidity. The larger the
$\sigma_0$ is, the larger the central pseudorapidity is and the
higher the peak is (, about $dE_T/d\eta\propto \sigma_0^{2\sim 3
}$ in region ($1.89 fm\leq\sigma_0\leq 3.26 fm$)). At the same
time, the initial energy density $\epsilon_0$ contribute the peak
value too (, $dE_T/d\eta\propto\epsilon_0$). But the location of
the peak is still invariant to $\epsilon_0$, due to our
assumption of ideal gas. To fix the location and peak value could
determine the initial parameters.

\begin{figure}[htb]
\begin{center}
\includegraphics[width=7.3cm]{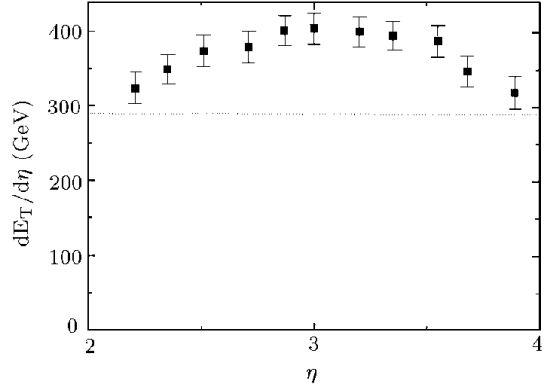}
\caption{The Pb+Pb data of NA49. The peak is at 3.0 unit. The
peak value is 405 $GeV$.}

\end{center}
\end{figure}

NA49 experiments showed\cite {sign:sign9} that there is a peak
around $\eta=3.0$, and the peak value is 405 $GeV$ per unit. (see
Fig 1.) Bjorken formula\cite {sign:sign10} gives $\epsilon_0=3.2$
$GeV/fm^3$ to fix the peak value. Setting $\epsilon_0=3.2$ $GeV$
and $\sigma_0=3.13$ $fm$ (both are near the upper limits) could
also fit the peak value well, (see Fig 2,) but the peak location
is $\eta=3.133$, a bit larger than expected and the peak width is
only about 1 unit far smaller than the experiment. The value
decrease quite rapidly far from the centre, while it is still
larger than 300 $GeV$ in experiment when $\eta=2$ and 4.

Here c.m. velocities of every small regions are used instead of
the particle velocities at freeze-out to compute the
pseudorapidity and transverse energy. Because this method did not
mention the fluid temperature and its components, one can not
calculate the real momentum distribution in that small region and
give the exact result. For this reason, the real curve should be
more smooth and the peak should be a little lower. It means the
real data are more far from the the experiment. Most of all, even
if the smooth effect dominate the distribution, the total $E_T$
and the peak location will not change much. Counted in Fig 2, the
total transverse energy between $\eta= 2.0$ to 4.0, is only about
433 $GeV$, while in Fig 1, the total transverse energy is no less
than 700 $GeV$. Even though set the effective size as 3.2 $fm$ to
the limit, the total transverse energy is only about 464 $GeV$,
while the peak value is 435.4 $GeV$ and its location comes up to
3.147 unit. Larger $\sigma_0$'s and initial energy densities than
those are not suitable.

So one reasonable explanation is that the experimental data\cite
{sign:sign9} contain a huge background\cite {sign:sign11}. This
background is probably produced by the other small fireballs if
multi-fireballs are emerged after the collision. The major signal
from the central fireball forms the peak and the others make up a
background.
By cutting off the background about 290$\pm$20 $GeV$\cite
{sign:sign11}, the peak value decreases to about 115$\pm$20 $GeV$
and total the transverse energy turns to 123$\pm$40 $GeV$.
Choosing a combination of $\sigma_0$ and $\epsilon_0$ properly,
could fit it very well.

\vspace{8pt}

\begin{center}
\begin{tabular}{|c|c|c|c|c|c|c|}
\hline $\epsilon_0$ $( GeV/fm^3 )$ & $1.60$ & $1.80$ & $2.00$ & $2.20$ & $2.40$  \\
\hline Peak value $( GeV / unit )$ & $89.3$ & $100.1$ & $111.8$ & $122.6$ & $133.9$   \\
\hline Sum $E_T$ $(GeV)$           & $104.0$ & $117.0$ & $130.0$ & $143.0$ & $156.0$  \\
\hline
\end{tabular}
\end{center}

\centerline{Table 1 \quad Possible $\epsilon_0$ at $\sigma_0=2.5$
$fm$ }

\vspace{8pt}
\quad

\begin{center}
\begin{tabular}{|c|c|c|c|c|c|c|}
\hline $\epsilon_0$ $( GeV/fm^3 )$ & $1.40$ & $1.60$ & $1.80$ & $2.00$ & $2.20$ \\
\hline Peak value $( GeV / unit )$ & $90.5$ & $103.1$ & $116.3$ & $129.0$ & $142.1$ \\
\hline Sum $E_T$ $(GeV)$          & $103.9$ & $118.8$ & $133.6$ & $148.5$ & $163.3$ \\
\hline
\end{tabular}
\end{center}

\centerline{Table 2 \quad Possible $\epsilon_0$ at $\sigma_0=2.6$
$fm$ }

\vspace{8pt}
\quad

\begin{figure}[htb]
  \centering
  \hspace{0.025\textwidth}%
  \begin{minipage}[t]{0.45\textwidth}
    \centering
    \includegraphics[width=\textwidth]{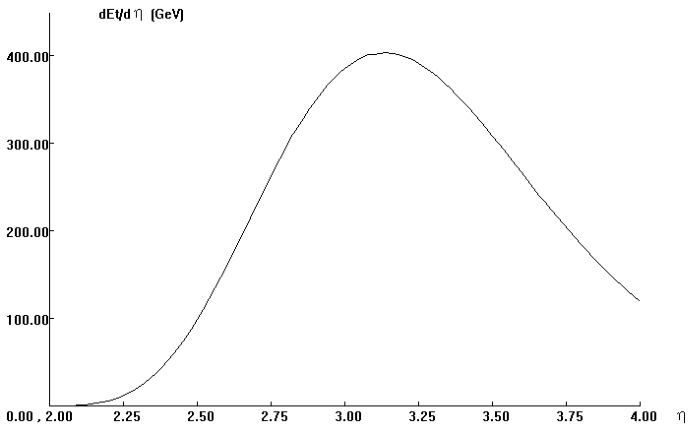}
    \caption{The $dE_T/d\eta$ distribution at $\epsilon_0=3.2$
$GeV/fm^3$ and $\sigma_0=3.13$ $fm$. The peak value is 403.9 $GeV$
and the peak location is about 3.13 unit.}
  \end{minipage}%
  \hspace{0.05\textwidth}%
  \begin{minipage}[t]{0.45\textwidth}
    \centering
    \includegraphics[width=\textwidth]{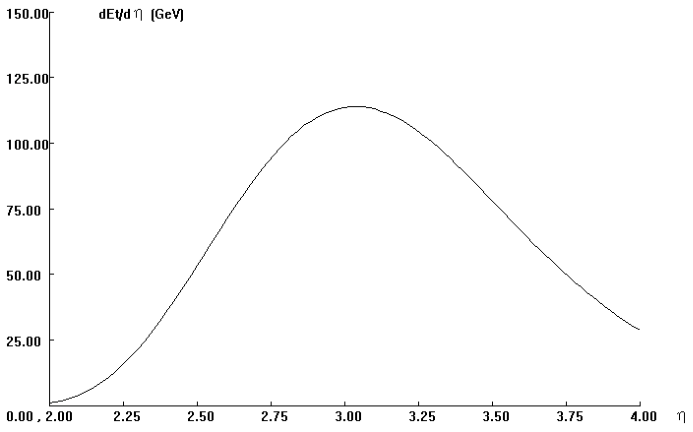}
    \caption{The $dE_T/d\eta$ distribution at $\epsilon_0=1.9$
$GeV/fm^3$ and $\sigma_0=2.55$ $fm$. The peak value is 114.0 $GeV$
and the peak location is about 3.03 unit.}
  \end{minipage}%
  \hspace{0.025\textwidth}
\end{figure}

Calculations show that from $\epsilon_0$ near 2.7 $GeV/fm^3$ at
$\sigma_0=2.3$ $fm$ ($\sigma_{freeze-out}=4.300$ $fm$),
$\epsilon_0$ near 2.0 $GeV/fm^3$ at $\sigma_0=2.5$ $fm$
($\sigma_{freeze-out}=4.350$ $fm$) to $\epsilon_0$ near 1.8
$GeV/fm^3$ at $\sigma_0=2.6$ $fm$ ($\sigma_{freeze-out}=4.375$
$fm$) are all permissive. Considering that the peak location
could not be too much larger or smaller than 3.0 and the
transverse velocity should not be too high, $\sigma_0$ is defined
to $[2.30,2.60]$.

The effective radius at freeze-out in experiment varies from 3.8
$fm$ by proton correlation\cite {sign:sign1}\cite {sign:sign12}
to $6.5\pm 0.5$ $fm$ by pion correlation\cite {sign:sign8}\cite
{sign:sign12}. This is quite dramatic and makes it pretty
difficult to determine the initial size precisely. The effective
radius got at freeze-out here varies from 4.15 $fm$ to 4.7 $fm$
according to the initial domain $\sigma_0=1.9\sim 3.2$ $fm$ (,
see Fig 4). This result is a little larger than the the data from
proton correlations and smaller than the data from pion
correlations. All could be acceptable, including our conclusion
$\sigma=4.3\sim4.4$ $fm$ which is quite near to the data from
proton correlations.

\begin{figure}[htb]
\begin{center}
\includegraphics[width=7.5cm]{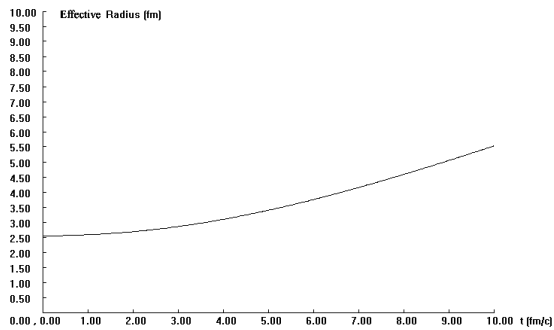}

\caption{Fireball size during evolution. $\epsilon_0=$1.9
$GeV/fm^3$, $\sigma_0=$2.55 $fm$.}

\end{center}
\end{figure}

Transverse velocity $v_T$= 0.55 is reported\cite
{sign:sign8}\cite {sign:sign13}. Again because the local thermal
momentum distribution and the particle component can not be
provided from the hydronynamic method, the c.m. velocity at the
effective radius $v_{\sigma}$ at freeze-out is used to compare
with the transverse velocity. From $\sigma_0=2.3$ $fm$ to 2.6
$fm$, $v_{\sigma}$ reduces from 0.645 to 0.613. We intend to
select a relatively larger initial effective radius with a lower
freeze-out velocity.

Energy density at freeze-out is estimated about 0.05
$GeV/fm^3$\cite {sign:sign1}. Evolution results with parameters
discussed above are about 0.051 to 0.065 $GeV/fm^3$, very
approximate. Smaller and larger $\sigma_0$ will produce too tiny
or huge result, although weone can tune the initial energy density
to give a small correction.

From above, our selected estimation of the initial parameters is
about $1.9\pm0.3$ $GeV/fm^3$ at $\sigma_0=2.55$ $fm$. Detailed
results are listed in Fig 5, 6, 7, 8. From Fig 5, one can see that
central energy density reduces very slow (about 0.1 $GeV/fm^3$) in
the first $1\sim2$ $fm/c$. State that the possibilities of
initial energy from range 1.4 to 2.4 $GeV/fm^3$ with relevant
effective size could not be removed completely either.

\begin{figure}[htb]
  \centering
  \hspace{0.025\textwidth}%
  \begin{minipage}[t]{0.45\textwidth}
    \centering
    \includegraphics[width=\textwidth]{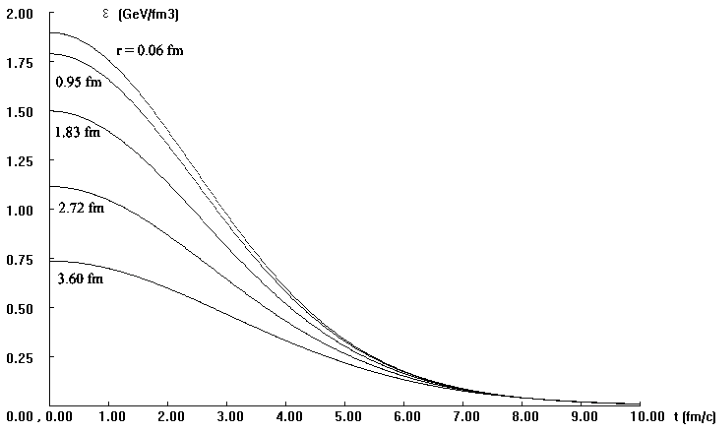}
    \caption{Energy density evolution at different locations.
$\epsilon_0=1.9$ $GeV/fm^3$, $\sigma_0=2.55$ $fm$.}
  \end{minipage}%
  \hspace{0.05\textwidth}%
  \begin{minipage}[t]{0.45\textwidth}
    \centering
    \includegraphics[width=\textwidth]{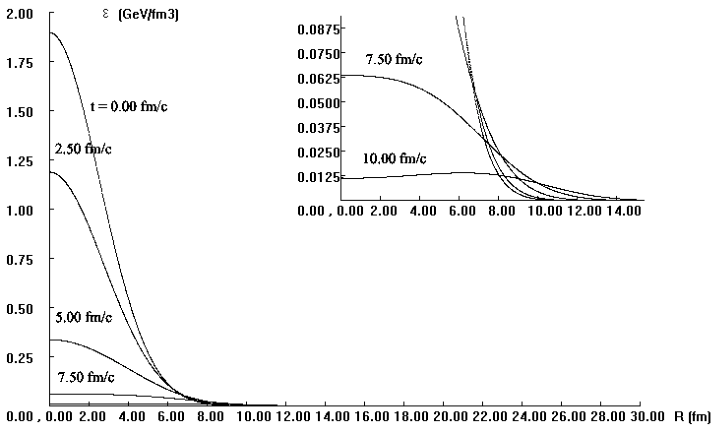}
    \caption{Energy density distributions at different time. $\epsilon_0=1.9$ $GeV/fm^3$, $\sigma_0=2.55$ $fm$.}
  \end{minipage}%
  \hspace{0.025\textwidth}
\end{figure}

\vspace{8pt}

\begin{figure}[hb]
  \centering
  \hspace{0.025\textwidth}%
  \begin{minipage}[t]{0.45\textwidth}
    \centering
    \includegraphics[width=\textwidth]{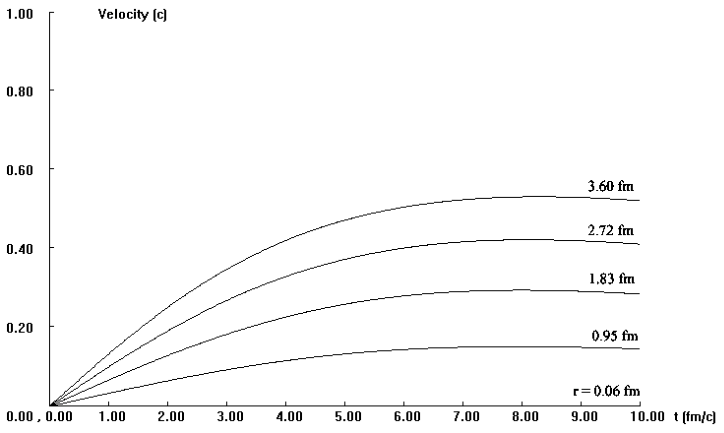}
    \caption{Fluid velocity evolution at different locations. The
velocities are from a set locations, but not according to the same
fluid parts. $\epsilon_0=1.9$ $GeV/fm^3$, $\sigma_0=2.55$ $fm$.}
  \end{minipage}%
  \hspace{0.05\textwidth}%
  \begin{minipage}[t]{0.45\textwidth}
    \centering
    \includegraphics[width=\textwidth]{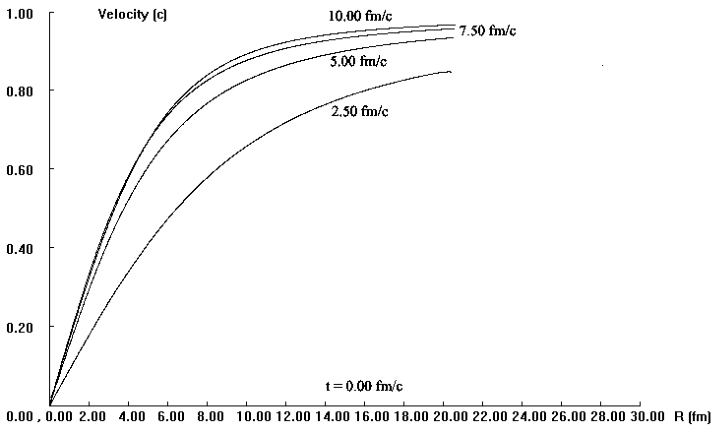}
    \caption{Fluid velocity distributions at different time. $\epsilon_0=1.9$ $GeV/fm^3$, $\sigma_0=2.55$ $fm$.}
  \end{minipage}%
  \hspace{0.025\textwidth}
\end{figure}

\vspace{12pt}

\leftline{ \bf 5. Cylindrical Fireball }

\vspace{8pt}

In this chapter, a short discuss is done on the evolution of a
flat cylindrical fireball. One more equation is added to the
evolution equations (10) and one dimension of data in memory
expand too. The hydrodynamic equation(s) could be written to these
forms

\begin{eqnarray}
\nonumber \left[\frac{\partial u_r ^2}{\partial r}+\frac{\partial
(u_ru_z)}{\partial z}+\frac{\partial(u_ru_0)}{\partial
t}\right]+(u_r ^2+\frac{1}{4})\frac{\partial\ln
\epsilon}{\partial r}+u_ru_z\frac{\partial\ln \epsilon}{\partial
z}+u_ru_0\frac{\partial\ln \epsilon}{\partial t}&+&\frac{u_r ^2}{r}=0 ,\\
\left[\frac{\partial(u_zu_r)}{\partial r}+\frac{\partial u_z
^2}{\partial z}+\frac{\partial(u_zu_0)}{\partial
t}\right]+u_zu_r\frac{\partial\ln \epsilon}{\partial r}+(u_z
^2+\frac{1}{4})\frac{\partial\ln \epsilon}{\partial
z}+u_zu_0\frac{\partial\ln \epsilon}{\partial t}&+&\frac{u_zu_r}{r}=0 ,\\
\nonumber \left[\frac{\partial(u_0u_r)}{\partial
r}+\frac{\partial (u_0u_z)}{\partial z}+\frac{\partial u_0
^2}{\partial t}\right]+u_0u_r\frac{\partial\ln \epsilon}{\partial
r}+u_0u_z\frac{\partial\ln \epsilon}{\partial z}+(u_0
^2-\frac{1}{4})\frac{\partial\ln \epsilon}{\partial
t}&+&\frac{u_0u_r}{r}=0.
\end{eqnarray}

Let $v_r=u_r/u_0$, $v_z=u_z/u_0,$ the partial time forms
\begin{eqnarray}
\nonumber\frac{\partial\ln \epsilon}{\partial
t}&=&\frac{4\alpha}{M}\left[2v_rA_0+2VzB_0+(\alpha-2)C_0\right],\\
\frac{\partial v_z}{\partial t}&=&-\frac{\alpha}{M}\left[2\alpha
v_zv_rA_0+(2\alpha v_z ^2+M)B_0+(\alpha V_z(\alpha-2)-Mv_r)
C_0\right],\\
\nonumber\frac{\partial v_r}{\partial
t}&=&-\frac{\alpha}{M}\left[2\alpha v_rv_zB_0+(2\alpha v_r
^2+M)A_0+(\alpha V_r(\alpha-2)-Mv_z)C_0\right],
\end{eqnarray}
where
\begin{eqnarray*}
A_0&=&\left[\frac{\partial u_r ^2}{\partial r}+\frac{\partial
(u_ru_z)}{\partial z}+\frac{\partial(u_ru_0)}{\partial
t}\right]+\frac{u_r ^2}{r},\\
B_0&=&\left[\frac{\partial(u_zu_r)}{\partial r}+\frac{\partial u_z
^2}{\partial z}+\frac{\partial(u_zu_0)}{\partial
t}\right]+\frac{u_zu_r}{r},\\
C_0&=&\left[\frac{\partial(u_0u_r)}{\partial r}+\frac{\partial
(u_0u_z)}{\partial z}+\frac{\partial u_0 ^2}{\partial
t}\right]+\frac{u_0u_r}{r},
\end{eqnarray*}
and
\begin{eqnarray*}
\alpha&=&1-v^2 _r-v^2 _z ,\\
M&=&\alpha ^2-2\alpha+4.
\end{eqnarray*}

\vspace{12pt}

\leftline{ \bf 6. Summary }

\vspace{8pt}

Hydrodynamic equation is used to compute a spherical fireball
created by the relativistic heavy ion collisions. The evolution
works very well. It can produce kinds of data to compare with
those from experiments. While, although these equations do not
have any free parameters, but due to the complex, unknown and
severe uncertain initial conditions, only a qualitative process
could be given. The estimate of initial data is only a kind of
attempt.

The experimental data is likely to contain a huge background. It
is reported that the initial energy density could be reduced to
0.91 $GeV/fm^3$\cite {sign:sign11}, by cutting off the
background. To use hydrodynamic method to deal with this problem
here, the initial energy density is estimated about
$\epsilon_0\approx1.9\pm0.3$ $GeV/fm^3$. Thinking that the
results are more sensitive to the initial size than to the
initial energy density, the real error range may be larger. The
result is not so striking as the estimation of $\epsilon_0=3.2$
$GeV/fm^3$ got before. The possibility of the QGP production in
CERN SPS is still not clear.

\vspace{12pt}

\leftline{ \bf Acknowledgement }

\vspace{8pt}

We would like to thank Professor ZHUANG Pengfei for the helpful
suggestions and discussions. This work was supported in part by
the National Natural Science Foundation of China (90103019), and
the Doctoral Programme Foundation of Institution of Higher
Education, the State Education Commission of China (2000000147).


\end{document}